\documentclass[sigconf, screen]{acmart}

\usepackage{booktabs}
\usepackage{float}
\usepackage{tabularx}
\usepackage{graphicx}
\usepackage{array}
\usepackage{threeparttable}
\usepackage{pifont}
\usepackage{multirow}
\usepackage{multicol}
\usepackage{makecell}

\newcommand{\cmark}{\ding{51}}%
\newcommand{\xmark}{\ding{55}}%

\title{Towards Personalized Bangla Book Recommendation: A Large-Scale Heterogeneous Book Graph Dataset}

\author{Rahin Arefin Ahmed}
\authornote{Both authors contributed equally to this research.}
\email{rahin520@gmail.com}
\author{Md. Anik Chowdhury}
\authornotemark[1]
\email{chowdhury.anik2000@gmail.com}
\affiliation{%
  \institution{East West University}
  \city{Dhaka}
  \country{Bangladesh}
}

\author{Sakil Ahmed Sheikh Reza}
\affiliation{%
  \institution{East West University}
  \city{Dhaka}
  \country{Bangladesh}}
\email{sakilahmedsheikhreza@gmail.com}

\author{Devnil Bhattacharjee}
\affiliation{%
  \institution{East West University}
  \city{Dhaka}
  \country{Bangladesh}}
\email{devnilbhattachargee@gmail.com}

\author{Muhammad Abdullah Adnan}
\affiliation{%
  \institution{Bangladesh University of Engineering and Technology}
  \city{Dhaka}
  \country{Bangladesh}}
\email{adnan@cse.buet.ac.bd}

\author{Julian McAuley}
\affiliation{%
  \institution{University of California San Diego}
  \city{San Diego, CA}
  \country{USA}}
\email{jmcauley@ucsd.edu}

\author{Nafis Sadeq}
\affiliation{%
  \institution{East West University}
  \city{Dhaka}
  \country{Bangladesh}}
\email{nafis.sadeq@ewubd.edu}

\renewcommand{\shortauthors}{Ahmed et al. (2026)}

\setcopyright{acmcopyright}
\acmConference{...}
\acmBooktitle{...}
\acmPrice{...}
\acmDOI{...}
\acmISBN{...}


\renewcommand\footnotetextcopyrightpermission[1]{}

\begin{document}

\begin{abstract}
Personalized book recommendation in Bangla literature has been constrained by the lack of structured, large-scale, and publicly available datasets. This work introduces RokomariBG, a large-scale heterogeneous book graph dataset designed to support research on personalized recommendation in a low-resource language setting. The dataset comprises 127,302 books, 63,723 users, 16,601 authors, 1,515 categories, 2,757 publishers, and 209,602 reviews, connected through several relation types and organized as a comprehensive knowledge graph. To demonstrate the utility of the dataset, we present a systematic benchmarking study on the top-N recommendation and sequential recommendation tasks, evaluating a diverse set of representative recommendation models. Through comprehensive benchmarking, we demonstrate that recommendation performance in this domain is strongly influenced by both heterogeneous relational information and code-mixed textual metadata. These findings reveal unique challenges of Bangladeshi e-commerce ecosystems that are largely absent from existing recommendation benchmarks. Overall, this work establishes a foundational benchmark and a publicly available resource for Bangla book recommendation research, enabling reproducible evaluation and future studies on recommendation in low-resource cultural domains. The dataset and code are publicly available at \url{https://github.com/backlashblitz/Bangla-Book-Recommendation-Dataset}
\end{abstract}

\maketitle

\section{Introduction}
Recommendation systems play a central role in modern online platforms by leveraging user preferences to personalize content discovery and improve user engagement~\cite{ricci2010introduction,bobadilla2013recommender}. Over the past decade, advances in recommendation research have been largely driven by the availability of large-scale, publicly accessible datasets, enabling systematic benchmarking and reproducible evaluation. Prominent examples include Amazon product reviews~\cite{mcauley2015image,he2016ups}, MovieLens~\cite{harper2015movielens}, and Goodreads~\cite{wan2018item}. Bangladesh represents a rapidly growing USD 7–10 billion e-commerce ecosystem serving nearly 180 million consumers~\cite{bangladeshi-ecommerce}, yet remains largely absent from mainstream recommender-system benchmarks despite its unique multilingual and code-mixed user behavior.

\begin{table*}
\caption{Dataset feature comparison in the Bangladeshi e-commerce domain}
\label{tab:dataset_comp}
\small
\resizebox{0.9\textwidth}{!}{
\begin{tabular}{p{4cm}|p{3cm}p{3.2cm}p{5cm}}
\toprule
Dataset & \multicolumn{3}{c}{Features} \\ \midrule
 & User-Item Interaction & Relational Knowledge & Side Features \\ \midrule
Rashid et al. (2024)~\cite{rashid2024comprehensive} & \xmark & \xmark & review, product name, category, rating \\
Ali et al. (2020)~\cite{ali2020banglasenti} & \xmark & \xmark & lexical sentiment polarity \\
Shamael et al. (2024)~\cite{shamael2024banglishrev} & \cmark & \xmark & review, seller reply, rating, likes, dislikes, dates, review image link \\
Shanto et al. (2023)~\cite{shanto2023mining} & \xmark & \xmark & comment, product type \\
Hossain et al. (2021)~\cite{hossain2021rating} & \cmark & \xmark & review, rating \\
Shanto et al.~\cite{shanto2023binary} & \xmark & \xmark & review text \\
Akter et al.(2021)~\cite{akter2021bengali}  & \cmark & \xmark & review , rating \\
Sarowar et al.{[}2019{]}~\cite{sarowar2019automated} & \xmark & \xmark & review \\
Bangla-Book-Review-Dataset  {[}2022{]}~\cite{sarker2022book} & \xmark & \xmark & review \\
Shafin et al. {[}2020{]}~\cite{shafin2020product} & \xmark & \xmark & review \\
Hossain et al. (2022)~\cite{hossain2022sentiment} & \xmark & \xmark & review \\
\textbf{RokomariBG} (Ours) & \cmark & Full book graph with eight relation types & 23 side features across five entities (Book, Author, Category, Publisher, Review) (Table~\ref{tab:entity-stats}) \\ \bottomrule
\end{tabular}
}
\end{table*}

\begin{table*}
\centering
\small
\caption{Entity statistics and attributes in \textbf{RokomariBG}}
\label{tab:entity-stats}
\begin{tabular}{lrl}
\toprule
\textbf{Entity Type} & \textbf{Count} & \multicolumn{1}{c}{\textbf{Key Attributes}} \\
\midrule
Books      & 127{,}302 & Book ID, Book Title, Summary, ISBN, Average Rating, Rating Count, Review Count, Pages \\
Authors    & 16{,}601  & Author ID, Author Name, Biography, Follower Count \\
Categories & 1{,}515   & Category ID, Category Name, Category Description, Total Book Count \\
Publishers & 2{,}757   & Publisher ID, Publisher Name, Publisher Description, Total Author Count, Total Book Count \\
Reviews    & 209{,}602 & Review ID, User Rating, Review Detail, Review Date, Up-vote, Down-vote, Verified Purchase \\
\bottomrule
\end{tabular}
\end{table*}

Research in the Bangla e-commerce domain has been predominantly driven by sentiment analysis. Existing datasets~\cite{rashid2024comprehensive, shamael2024banglishrev, akter2021bengali, shanto2023mining,hossain2021rating} primarily focus on review classification and lack essential user-item interactions required for personalized recommendation. As shown in Table~\ref{tab:dataset_comp}, most prior Bangladeshi datasets do not support recommendation research, as they miss user-item interactions and relational knowledge. In contrast, modern recommender systems have advanced significantly with collaborative filtering~\cite{hu2008collaborative}, matrix factorization~\cite{koren2009matrix}, graph neural networks~\cite{he2020lightgcn,zhang2019heterogeneous,wang2019kgat}, and neural collaborative filtering~\cite{covington2016deep,yi2019sampling, yang2020mixed}. However, these advancements remain largely unexplored in the Bangladeshi e-commerce domain due to the lack of suitable datasets.

Modern recommendation models can be broadly categorized based on the types of signals they exploit from available data. 
The first category consists of methods that rely primarily on user–item interaction signals, such as ratings, clicks, or purchase histories. 
Classical collaborative filtering (CF) and matrix factorization (MF) approaches fall into this category, including neighborhood-based methods and latent factor models that learn user and item representations solely from interaction matrices~\cite{sarwar2001item,koren2009matrix,rendle2009bpr,hu2008collaborative}. 

The second category extends interaction modeling by incorporating relational knowledge among multiple entities. Graph-based recommendation models represent users, items, and related entities as nodes in a graph and leverage message passing to propagate signals across relational structures. Representative examples include LightGCN, which simplifies graph convolution to improve scalability and performance in collaborative settings~\cite{he2020lightgcn}, and heterogeneous graph neural networks (HGNNs), which explicitly model multiple node and edge types through relation-aware or meta-path–based message propagation~\cite{zhang2019heterogeneous}. These models have demonstrated strong performance gains by capturing higher-order connectivity and structural dependencies beyond direct user–item interactions. 

The third category further enriches recommendation models by jointly modeling interaction signals, relational knowledge, and natural language content. Such approaches integrate textual features from item descriptions, reviews, or user-generated content alongside graph or interaction-based representations, enabling semantic understanding in addition to structural reasoning~\cite{zhang2016cke,wang2018dkn}. Despite the effectiveness of these advanced modeling paradigms, their application to Bangla book recommendation has remained largely unexplored due to the lack of large-scale, structured, and multi-entity benchmark datasets. 

To bridge this critical gap, this work makes three contributions. First, we construct and release RokomariBG, a large-scale heterogeneous book graph dataset constructed from Rokomari.com, Bangladesh’s largest online bookstore. The dataset comprises 127,302 book entities, 16,601 authors, 1,515 categories, 2,757 publishers, 63,723 anonymized users, and 209,602 user-generated reviews. These entities are connected through eight distinct relationship types, forming a rich heterogeneous knowledge graph that supports advanced recommendation modeling and multi-relational learning.

Second, to establish a strong empirical foundation for future research, we provide a comprehensive benchmarking study on the top-N recommendation and sequential tasks. We evaluate a broad spectrum of representative recommendation approaches, including popularity-based baselines, collaborative filtering methods, matrix factorization models, content-based approaches, hybrid models with side information, neural two-tower retrieval architectures, and several sequential recommendation models. Our experiment results underscore the importance of jointly modeling user–item interactions, relational structure, and auxiliary textual signals in low-resource recommendation scenarios.

Third, we conduct a detailed exploratory data analysis to characterize user behavior and content consumption patterns in Bangla literature. The analysis reveals several distinctive properties of the ecosystem, including a high concentration of five-star ratings (65.8\%) and strong author-loyalty exhibited by Bangladeshi readers. These observations provide practical insights for platform operators and content creators, while also informing the design of recommendation models tailored to Bangla reading habits.

Overall, by releasing this dataset and accompanying benchmark, this work establishes a foundational resource for Bangla book recommendation research. Beyond enabling reproducible evaluation in a low-resource language, the dataset offers opportunities for future studies on heterogeneous graph learning, representation learning with sparse interactions, and cross-lingual or multilingual recommendation systems.

\section{Dataset}
The dataset statistics can be found in Table~\ref{tab:entity-stats} and Table~\ref{tab:relations}. The largest entity type is \emph{Books}, with \textbf{127,302} unique items. Each book is associated with structured metadata such as ISBN, number of pages, average rating, rating count, and review count, as well as unstructured textual content including the book title and summary. These attributes enable both collaborative and content-based modeling.

The dataset includes \textbf{16,601} \emph{Authors}, each represented by an author ID, name, biographical text, and follower count, allowing the modeling of author popularity and author-level semantics. Books are also linked to \textbf{2,757} \emph{Publishers}, where each publisher entity contains descriptive text and aggregate statistics such as total book count and total author count. In addition, \textbf{1,515} \emph{Categories} are provided with each category associated with a name, description, and the total number of books under that category.

User feedback is captured through \textbf{209,602} \emph{Reviews}. Each review includes a user-provided rating, detailed review text, timestamp, up-vote and down-vote counts, and a verified-purchase indicator. Reviews serve as the primary interaction signal in the dataset and connect users to books through explicit feedback.

The relational structure of the dataset contains the following relation types: Book–Author \textbf{(97,497)}, Book– Publisher \textbf{(94,957)}, Book-
Category \textbf{(235,496)}, Author–Category \textbf{(98,305)}, Author–Publisher \textbf{(40,630)}, Publisher-Category \textbf{(18,802)} edges. There are \textbf{209,602} review connections to represent user–item interactions.

\begin{table}
\caption{Relationship statistics.}
\label{tab:relations}
\small
\begin{tabular}{lrl}
\toprule
\textbf{Relation} & \textbf{Count} & \textbf{Relation Type} \\
\midrule
Book $\leftrightarrow$ Author & 97,497 & Many-to-Many \\
Book $\leftrightarrow$ Publisher & 94,957 & Many-to-One \\
Book $\leftrightarrow$ Category & 235,496 & Many-to-Many \\
Author $\leftrightarrow$ Category & 98,305 & Many-to-Many \\
Author $\leftrightarrow$ Publisher & 40,630 & Many-to-Many \\
Publisher $\leftrightarrow$ Category & 18,802 & Many-to-Many \\
Review $\leftrightarrow$ User & 209,602 & Many-to-One \\
Review $\leftrightarrow$ Book & 209,602 & Many-to-One \\
\bottomrule
\end{tabular}
\end{table}

\subsection{Dataset Construction}

\paragraph{\textbf{Entity and Metadata Extraction}}
The dataset was constructed by crawling publicly accessible reviews, entity relations, and metadata from \textit{Rokomari.com}. Only publicly available information was collected, following the website’s \texttt{robots.txt}. \textit{BeautifulSoup} was used for HTML parsing, and raw HTML files were systematically transformed into structured JSON representations. Extracted entities—including books, categories, authors, publishers, and reviews— were stored as individual JSON files following a standardized schema.

\paragraph{\textbf{Entity Linking}}
Rokomari webpages associated with core entities such as books, authors, categories, and publishers contain hyperlinks to the corresponding webpages of related entities. Each hyperlink is unique and embeds an entity-specific identifier within the URL. Relational edges were constructed by resolving these hyperlink references embedded in the HTML pages, enabling the reconstruction of entity relationships.

\paragraph{\textbf{Deduplication and pre-processing}}
During the web crawling process, duplicate HTML files were detected using URL hashing and entity identifier matching. 
Data normalization involved parsing comma-separated numeric values (e.g., ``1,250'' to 1250), constraining rating values to the $[1.0, 5.0]$ range, and standardizing price formats, Bangla numeric notation, and rating fields. Missing attributes were explicitly represented as \texttt{null} values. Referential integrity was preserved by validating all relational edges against the corresponding entity identifier sets. %

\paragraph{\textbf{Privacy Preservation}}
To protect user privacy, all personally identifiable information (PII) was removed from the dataset. Users were anonymized using privacy-preserving sequential identifiers that do not retain usernames, profile images, email addresses, or any other identifying information.

\subsection{Data Quality Analysis}

\paragraph{\textbf{Metadata Completeness}} As shown in Table~\ref{tab:metadata-completeness}, metadata completeness is evaluated for 127,302 unique books. The dataset provides strong coverage of key attributes: book title achieves 100\% fully, while book summary, price, rating and review count each achieve 99.7\%. Relational attributes such as categories (84.6\%) and publishers (74.6\%) are sufficiently represented to support multi-relational graph analysis. The ISBN completeness rate is 63.4\%.

\paragraph{\textbf{Review Quality}} Review quality indicators are shown in Table~\ref{tab:review-quality}. The dataset contains 209,602 reviews, the majority of which are content rich: 83.3\% feature more than 10 characters of text, offering valuable signals for personalized preference and sentiment analysis. Explicit user ratings on a 1--5 star scale are available in 82.6\% of reviews, providing support for both explicit and implicit feedback models. Verified purchase status is available in 55.0\% of reviews and 41.6\% of reviews have at least one helpfulness vote (positive or negative). These annotations can be used as quality metrics for review weighting and trust evaluation.

\begin{table}
\small
    \centering
    \caption{Metadata completeness for core book attributes.}
    \label{tab:metadata-completeness}
    \begin{tabular}{lrr}
    \toprule
    \textbf{Metadata Field} & \textbf{Count} & \textbf{Completeness (\%)} \\
    \midrule
    Title           & 127,302 & 100.0 \\
    Book ID         & 127,302 & 100.0 \\
    Category        & 107,697 &  84.6 \\
    Publisher       &  94,957 &  74.6 \\
    Rating          & 126,955 &  99.7 \\
    Review Count    & 126,966 &  99.7 \\
    Pages           & 100,781 &  79.2 \\
    ISBN            &  80,739 &  63.4 \\
    Summary         & 126,966 &  99.7 \\
    \bottomrule
    \end{tabular}
\end{table}

\begin{table}
\centering
\caption{Review Quality Indicators}
\label{tab:review-quality}
\small
\begin{tabular}{lrr}
\toprule
\textbf{Quality Metric}          & \textbf{Count} & \textbf{Percentage (\%)} \\
\midrule
Verified Purchase                & 115,210        & 55.0                     \\
Has Rating                       & 173,111        & 82.6                     \\
Has Text ($>10$ chars)           & 174,651        & 83.3                     \\
Has Votes                        & 87,234         & 41.6                     \\
\bottomrule
\end{tabular}
\end{table}

\paragraph{\textbf{Interaction Sparsity Analysis}}
User–item interactions in the RokomariBG dataset exhibit substantial sparsity on both the item and user sides, a characteristic commonly observed in real-world e-commerce and review platforms. Table~\ref{tab:book-engagement-distribution} summarizes the distribution of review engagement across books. While a significant portion of books receive moderate to high attention, interaction coverage remains uneven. Specifically, \textbf{5.53\%} of books have no reviews at all, and an additional \textbf{28.23\%} of books receive at most two reviews, indicating limited exposure for a non-trivial fraction of the catalog. At the same time, \textbf{42.57\%} of books receive five or more reviews, suggesting a popularity-driven concentration of user attention on a relatively small subset of items.

User activity is even more skewed, as shown in Table~\ref{tab:user-activity-distribution}. User engagement follows a pronounced long-tail (power-law) distribution: a majority of users (\textbf{53.7\%}) contribute only a single review, and \textbf{84.5\%} of users provide fewer than five reviews in total, reflecting predominantly casual participation. In contrast, highly active users are extremely rare—only \textbf{1.6\%} of users contribute between 20 and 49 reviews, and just \textbf{0.4\%} qualify as power users with 50 or more reviews.

Additional details regarding the properties of the dataset such as top authors and categories by user engagement, rating distribution, linguistic composition of review texts, and cold start statistics for each category can be found in Appendix~\ref{app:aeda}.

\begin{table}
\centering
\caption{Book interaction sparsity}
\label{tab:book-engagement-distribution}
\small
\begin{tabular}{lrr}
\toprule
\textbf{Engagement Category} & \textbf{Book Count} & \textbf{Percentage (\%)} \\
\midrule
0 reviews  & 1,338 & 5.53 \\
1 review   & 243 & 1.00 \\
2 reviews  & 6,592 & 27.23 \\
3 reviews  & 3,555 & 14.69 \\
4 reviews  & 2,175 & 8.98 \\
5 or more reviews & 10,305 & 42.57 \\
\midrule
\textbf{Total} & \textbf{24,208}  \\
\bottomrule
\end{tabular}
\end{table}

\begin{table}
\centering
\caption{User interaction sparsity}
\label{tab:user-activity-distribution}
\small
\begin{tabular}{lrr}
\toprule
\textbf{Review Activity} & \textbf{\# Users} & \textbf{Percentage (\%)} \\
\midrule
1 review                 & 34,240  & 53.7 \\
2--4 reviews             & 19,651  & 30.8 \\
5--9 reviews             & 6,075   & 9.5 \\
10--19 reviews           & 2,495   & 3.9 \\
20--49 reviews           & 1,027   & 1.6 \\
50 or more reviews       & 235     & 0.4 \\
\midrule
\textbf{Total}           & \textbf{63,723}  \\
\bottomrule
\end{tabular}
\end{table}

\section{Methodology}

\label{sec:methodology}

For benchmarking on the proposed Bangla book graph dataset, we implement a hybrid neural two-tower (dual-encoder) recommendation architecture as an advanced baseline. This architecture provides a strong and extensible neural baseline for evaluating the impact of interaction signals, side features, and multi-entity relational knowledge in Bangla book recommendation. Two-tower models are particularly well-suited for large-scale Top-$N$ recommendation, as they enable efficient candidate retrieval by embedding users and items independently into a shared latent space and scoring them via vector similarity. 

The model consists of a \emph{user tower} and an \emph{item tower}, each parameterized by a multi-layer perceptron (MLP). Given a user $u$ and an item $i$, the model produces normalized embeddings $\mathbf{z}_u, \mathbf{z}_i \in \mathbb{R}^d$, and their relevance is computed using the dot product:
\[
s(u,i) = \mathbf{z}_u^\top \mathbf{z}_i.
\]
During training, the model is optimized to rank observed user--item interactions higher than non-interacted items using in-batch negatives.

\subsection{Item Tower}

The item tower encodes a book by jointly modeling three complementary sources of information: (i) item identity, (ii) side features, and (iii) relational knowledge derived from the heterogeneous book graph.

\paragraph{Item Identity.}
Each book is associated with a trainable item ID embedding $\mathbf{e}_i \in \mathbb{R}^{d_{\text{id}}}$, which captures collaborative filtering signals from user--item interactions.

\paragraph{Side Features.}
Side features consist of both textual and numeric metadata:
\begin{itemize}
    \item \textbf{Textual features:} For each book, a textual description is constructed by concatenating the title, summary, author information, category descriptions, and publisher metadata. This text is encoded using a off-the-shelf text embedding model, producing a dense text embedding $\mathbf{t}_i$. A linear projection layer maps $\mathbf{t}_i$ to a lower-dimensional space compatible with the item representation.
    \item \textbf{Numeric features:} Structured numeric attributes such as price, number of pages, rating statistics, and popularity indicators are normalized and concatenated as a dense feature vector $\mathbf{n}_i$.
\end{itemize}

\paragraph{Relational Knowledge.}
To leverage the multi-entity structure of the dataset, we incorporate relational information from authors, categories, and publishers. Each entity type is represented by its own embedding table. For books linked to multiple authors or categories, the corresponding entity embeddings are mean-pooled:
\[
\mathbf{a}_i = \frac{1}{|\mathcal{A}_i|} \sum_{a \in \mathcal{A}_i} \mathbf{e}_a, \quad
\mathbf{c}_i = \frac{1}{|\mathcal{C}_i|} \sum_{c \in \mathcal{C}_i} \mathbf{e}_c,
\]
where $\mathcal{A}_i$ and $\mathcal{C}_i$ denote the sets of authors and categories associated with book $i$, respectively. Publisher information is modeled using a single embedding $\mathbf{p}_i$.

\paragraph{Item Representation.}
Depending on the experimental setting, the item tower concatenates the available components:
\[
\mathbf{x}_i = \big[ \mathbf{e}_i \,;\, \mathbf{p}_i \,;\, \mathbf{a}_i \,;\, \mathbf{c}_i \,;\, \mathbf{n}_i \,;\, \mathbf{t}_i \big],
\]
where omitted components correspond to disabled signals in ablation experiments. The concatenated vector is passed through an MLP and $\ell_2$-normalized to obtain the final item embedding $\mathbf{z}_i$.

\subsection{User Tower}

The user tower encodes user preferences by combining a user ID embedding with a representation of recent interaction history.

\paragraph{User Identity.}
Each user $u$ is associated with a trainable embedding $\mathbf{e}_u \in \mathbb{R}^{d_{\text{id}}}$.

\paragraph{Interaction History Pooling.}
To capture short-term and long-term preferences, we construct a user history representation by averaging the item embeddings of the most recent $K$ interacted books:
\[
\mathbf{h}_u = \frac{1}{|\mathcal{H}_u|} \sum_{i \in \mathcal{H}_u} \mathbf{z}_i,
\]
where $\mathcal{H}_u$ denotes the set of historical items for user $u$ (truncated to a maximum length $K$). If no history is available, a zero vector is used.

\paragraph{User Representation.}
The final user input vector is formed as:
\[
\mathbf{x}_u = \big[ \mathbf{e}_u \,;\, \mathbf{h}_u \big],
\]
which is passed through a user MLP and normalized to produce the user embedding $\mathbf{z}_u$. In the interaction-ablation setting, the user representation is replaced by a shared global embedding, effectively removing personalization derived from interaction data.

\subsection{Training Objective}

The model is trained using a softmax cross-entropy loss with in-batch negatives. Given a mini-batch of $B$ positive user--item pairs, the score matrix $\mathbf{S} \in \mathbb{R}^{B \times B}$ is computed as:
\[
S_{jk} = \mathbf{z}_{u_j}^\top \mathbf{z}_{i_k}.
\]
The objective encourages each user embedding to assign the highest score to its corresponding positive item. Interaction weights derived from review metadata (e.g., verified purchase and rating strength) are used to reweight the loss. At inference time, all item embeddings are precomputed, enabling efficient Top-$N$ retrieval via nearest-neighbor search in the embedding space.

\begin{table*}[t]
\centering
\caption{Performance comparison of recommendation models.
Results are reported as mean $\pm$ standard deviation over 3 runs with different random seeds.
Best results are bolded.}
\label{tab:model-comparison}
\small
\begin{tabular}{lcccccc}
    \toprule
    \textbf{Model} & \textbf{Hit@5} & \textbf{Hit@10} & \textbf{Hit@50} & \textbf{MRR@10} & \textbf{NDCG@10} & \textbf{NDCG@50} \\
    \midrule
    Popularity 
        & 0.033 $\pm$ 0.001  & 0.073 $\pm$ 0.001  & 0.073 $\pm$ 0.001 
        & 0.021 $\pm$ 0.001  & 0.033 $\pm$ 0.001 & 0.033 $\pm$ 0.001  \\

    Category-Aware Popularity 
        & 0.136 $\pm$ 0.001 & 0.162 $\pm$ 0.002 & \textbf{0.348 $\pm$ 0.001} 
        & 0.117 $\pm$ 0.002 & \textbf{0.128 $\pm$ 0.002} & \textbf{0.167 $\pm$ 0.001} \\

    User-Based CF 
        & 0.042 $\pm$ 0.002 & 0.057 $\pm$ 0.002 & 0.057 $\pm$ 0.002 
        & 0.024 $\pm$ 0.001 & 0.032 $\pm$ 0.002 & 0.032 $\pm$ 0.002 \\

    Item-Based CF 
        & 0.087 $\pm$ 0.001 & 0.101 $\pm$ 0.000 & 0.101 $\pm$ 0.001 
        & 0.056 $\pm$ 0.001 & 0.067 $\pm$ 0.002 & 0.067 $\pm$ 0.001 \\

    Implicit MF 
        & 0.027 $\pm$ 0.001 & 0.040 $\pm$ 0.000 & 0.040 $\pm$ 0.001
        & 0.016 $\pm$ 0.001 & 0.022 $\pm$ 0.000 & 0.022 $\pm$ 0.001 \\

    Explicit MF 
        & 0.001 $\pm$ 0.000 & 0.003 $\pm$ 0.001 & 0.003 $\pm$ 0.000
        & 0.000 $\pm$ 0.000 & 0.001 $\pm$ 0.000 & 0.001 $\pm$ 0.000 \\

    Pure Content-Based 
        & \textbf{0.137 $\pm$ 0.000} & \textbf{0.172 $\pm$ 0.000} & 0.243 $\pm$ 0.000
        & \textbf{0.192 $\pm$ 0.001} & 0.108 $\pm$ 0.000 & 0.124 $\pm$ 0.000\\

    Hybrid: MF + Side 
        & 0.092 $\pm$ 0.000 & 0.122 $\pm$ 0.000 & 0.211 $\pm$ 0.000  
        & 0.050 $\pm$ 0.000 & 0.067 $\pm$ 0.000 & 0.086 $\pm$ 0.000 \\

    LightGCN\cite{he2020lightgcn} 
        & 0.088 $\pm$ 0.001 & 0.107 $\pm$ 0.001 & 0.162 $\pm$ 0.001 
        & 0.058 $\pm$ 0.001 & 0.070 $\pm$ 0.001 & 0.082 $\pm$ 0.002 \\

    HGNN + Side\cite{zhang2019heterogeneous} 
        & 0.063 $\pm$ 0.001 & 0.109 $\pm$ 0.001 & 0.185 $\pm$ 0.001 
        & 0.044 $\pm$ 0.002 & 0.075 $\pm$ 0.001 & 0.098 $\pm$ 0.002 \\

    Neural Two-Tower (NTT) + Side\cite{yi2019twotower} 
        & 0.133 $\pm$ 0.001 & \textbf{0.172 $\pm$ 0.002} & 0.312 $\pm$ 0.002 
        & 0.094 $\pm$ 0.002 & 0.113 $\pm$ 0.001 & 0.143 $\pm$ 0.001 \\

    Review-Augmented NTT (RA-NTT)
        & 0.122 $\pm$ 0.001 & 0.156 $\pm$ 0.002 & 0.300 $\pm$ 0.002 
        & 0.086 $\pm$ 0.002 & 0.103 $\pm$ 0.001 & 0.134 $\pm$ 0.001 \\    
    \bottomrule
\end{tabular}
\end{table*}

\section{Experiments}
\subsection{Experimental Setup}

\paragraph{\textbf{Dataset Split}} We construct a unified interaction table by joining user--review and book--review relations through review identifiers, resulting in $(u,i)$ interaction pairs. We use a per user temporal leave-last-one-out protocol for dataset split in top-N recommendation. The last interaction of every user is used as the test sample and second last interaction is used as the validation sample. The rest of the interactions are left in the training split. 

\subsection{Benchmarking Models}
\paragraph{\textbf{Top-N recommendation}} We evaluate a diverse set of recommendation models on the top-N task, including non-personalized baselines (Global Popularity and Category-Aware Popularity), classical collaborative filtering (User-CF and Item-CF with $  k=50  $), matrix factorization methods (Implicit ALS with 64 dimensions and Explicit SVD), a content-based model using multi-hot encodings and TF-IDF features, a hybrid MF model with side features (LightFM), graph-based models (LightGCN and HGNN with side information), and a Neural Two-Tower retrieval. The Neural Two-Tower model jointly encodes user and item representations using ID embeddings, interaction history, textual features (via multilingual sentence transformers), numeric metadata and multi-entity relational knowledge, trained with in-batch negative sampling. 

We further extend the baseline Neural Two-Tower model with a Review-Augmented variant (RA-NTT), which incorporates user-written review text into the user tower via a learned soft-gate mechanism. Each user's reviews are encoded into a 768-dimensional vector using a multilingual encoder~\cite{wang2024multilinguale5textembeddings} and concatenated with the interaction-based user representation.

\paragraph{\textbf{Sequential recommendation}} For sequential recommendation, we use simple baselines such recent-category-popularity, continue-author, item-to-item transition, personal recency-weighted popularity as well as neural models like GRU4Rec~\cite{Hidasi2015RECURRENTNN}, SASRec~\cite{Kang2018SelfAttentiveSR}, and BERT4Rec~\cite{Sun2019BERT4RecSR}.	

\subsection{Results and Discussion}

\paragraph{\textbf{Side Features Significantly Improve Ranking Quality}}
The strongest recommendation models consistently incorporate side information beyond user--item interactions. In terms of NDCG@10, the Pure Content-Based model achieves 0.108, substantially outperforming UserCF (0.032), ItemCF (0.067), Implicit MF (0.022), and Explicit MF (0.001). Similarly, Hybrid MF with side features improves upon matrix factorization baseline, increasing NDCG@10 from 0.022 to 0.067. These results suggest that metadata associated with Bangla books provides highly informative signals for recommendation.

\paragraph{\textbf{Relational Knowledge is Crucial for Bangla Books}}
Models that exploit relational structure achieve relatively improved ranking performance. HGNN, which models relationships among books and auxiliary entities through graph propagation, attains an NDCG@10 of 0.075, outperforming traditional collaborative filtering and matrix factorization approaches. The Neural Two-Tower model, which further incorporates multi-entity relational knowledge alongside textual and metadata features, achieves eve better performance with an NDCG@10 of 0.103. The consistent advantage of HGNN and the Neural Two-Tower architecture suggests that user preferences in the Bangla book domain are strongly influenced by relationships among authors, publishers, categories, and books.

\paragraph{\textbf{Strong Author Loyalty Among Bangladeshi Readers}}
The Continue Author baseline achieves the highest Hit@10 (0.163) and Hit@50 (0.251), outperforming considerably more sophisticated neural sequential recommenders, including GRU4Rec (0.070/0.140), SASRec (0.074/0.142), and BERT4Rec (0.078/0.136). The effectiveness of this simple heuristic suggests that readers frequently consume multiple books by the same author in succession, indicating a strong author-loyalty effect among Bangladeshi readers.

\begin{table}
  \caption{Results for sequential Recommendation}
  \label{tab:sequential-results}
  \centering
  \small
  \setlength{\tabcolsep}{3pt}
  \begin{tabular}{lccc}
    \toprule
    \textbf{Model} & \textbf{Hit@5} & \textbf{Hit@10} & \textbf{Hit@50} \\
    \midrule 
    Recent Category Pop.     & 0.079 & 0.105 & 0.174 \\
    Continue Author          & 0.134 & \textbf{0.163} & \textbf{0.251} \\
    Item-to-Item (last 3)    & 0.134 & 0.158 & 0.229 \\
    Personal Recency-Weighted& 0.067 & 0.097 & 0.217 \\
    GRU4Rec\cite{Hidasi2015RECURRENTNN}                   & 0.042 & 0.070 & 0.140 \\
    SASRec\cite{Kang2018SelfAttentiveSR}                   & 0.048 & 0.074 & 0.142 \\
    BERT4Rec\cite{Sun2019BERT4RecSR}                 & 0.060 & 0.078 & 0.136 \\
    \bottomrule
  \end{tabular}
\end{table}

\subsection{Ablation Study}
To quantify the contribution of different information sources, we perform ablation studies on the Neural Two-Tower model by removing side features (text embeddings and numeric metadata), multi-entity relational knowledge, and interaction signals. Results are summarized in Table~\ref{tab:ablation_ntt}.

\paragraph{\textbf{Warm-start.}} The full model achieves NDCG@10 = 0.103 and NDCG@50 = 0.133. Removing side features reduces NDCG@10 to 0.088, indicating that metadata provides useful complementary signals beyond interaction history. Eliminating relational knowledge causes a larger decline to 0.076, highlighting the importance of relationships among books, authors, publishers, and categories. The largest performance degradation occurs when interaction signals are removed, with NDCG@10 dropping to 0.028. These results confirm that collaborative signals remain the primary driver of recommendation quality when sufficient user history is available, while side information and relational knowledge provide substantial additional gains.

\paragraph{\textbf{Cold-start.}} The relative importance of information sources changes considerably in the cold-start setting. The full model achieves NDCG@10 = 0.055 and NDCG@50 = 0.070. Removing relational knowledge produces the largest degradation, reducing NDCG@10 to 0.027, followed by the removal of side features, which lowers performance to 0.034. In contrast, removing interaction signals results in a comparatively smaller decrease to 0.042. This shift indicates that auxiliary information becomes the dominant source of personalization when little or no user history is available. %

\begin{table}
\centering
\small
\caption{Ablation study on the two-tower retrieval model under warm and cold settings. Cold-start evaluation considers users with one or no interactions in the training set.}
\label{tab:ablation_ntt}
\begin{tabular}{l|lcc}
\toprule
\textbf{Scenario} & \textbf{Setting} & \textbf{NDCG@10} & \textbf{NDCG@50} \\
\midrule
\multirow{4}{*}{Warm-start}
 & Full model        & 0.103 & 0.133\\
 & -- Side features  & 0.088 & 0.125 \\
 & -- Relations      & 0.076 & 0.106 \\
 & -- Interaction    & 0.028 & 0.061 \\
\midrule
\multirow{4}{*}{Cold-start}
 & Full model        & 0.055 & 0.070 \\
 & -- Side features  & 0.034 & 0.053 \\
 & -- Relations      & 0.027 & 0.035 \\
 & -- Interaction    & 0.042 & 0.063 \\
\bottomrule
\end{tabular}
\end{table}

\begin{table}
\centering
\small
\caption{Performance of the two-tower retrieval model using different text embedding models.}
\label{tab:text_embedding}
\begin{tabular}{lcc}
\toprule
\textbf{Text Embedding Model} & \textbf{NDCG@10} & \textbf{NDCG@50} \\
\midrule
multilingual-e5-base~\cite{wang2024multilinguale5textembeddings}
    & 0.088 & 0.123 \\
bangla-sentence-transformer~\cite{udding-2024-bn-sbert}
    & 0.094 & 0.128 \\
paraphrase-multilingual-L12-v2~\cite{reimers-2019-sentence-bert}
    & 0.090 & 0.127 \\
multilingual-e5-large-instruct~\cite{wang2024multilinguale5textembeddings}
    & \textbf{0.103} & \textbf{0.133} \\
\bottomrule
\end{tabular}
\end{table}

\section{Related Works}

Research is Bangla e-commerce domain, has been largely driven by sentiment analysis. Both foundational and recent works have established strong baselines for classifying user opinions from product reviews, employing a wide range of techniques from traditional machine learning with handcrafted features~\cite{tabassum2019design} to advanced deep learning architectures~\cite{sarker2022book}. Major contributions include the development of sentiment lexicons~\cite{ali2020banglasenti}, methods for processing multilingual and code-mixed (Bangla--English) text~\cite{mukit2023sentiment, shamael2024banglishrev}, and the curation of dedicated domain-specific datasets for this task~\cite{shanto2023mining, rashid2024comprehensive}. Overall, these efforts have achieved high classification accuracy, demonstrating robust sentiment comprehension for Bangla content~\cite{akter2021bengali, zulfiker2022bangla, hossain2022sentiment}.

Meanwhile, the broader field of recommender systems has undergone a substantial shift from early collaborative filtering~\cite{sarwar2001item, koren2009matrix} and matrix factorization techniques~\cite{rendle2009bpr, hu2008collaborative} to neural and graph-based architectures. The two-tower neural network design~\cite{covington2016deep}, optimized through advanced negative sampling strategies~\cite{yi2019sampling, yang2020mixed}, has become a widely accepted approach for scalable retrieval. A more transformative advancement has been the integration of Graph Neural Networks (GNNs), which effectively model the relational structure of user–item interactions. Architectures like Neural Graph Collaborative Filtering (NGCF)~\cite{wang2019neural} and LightGCN~\cite{he2020lightgcn} capture high-order collaborative signals through graph convolutions. For complex, multi-typed data, HGNNs~\cite{zhang2019heterogeneous} and knowledge graph-enhanced models like KGAT~\cite{wang2019kgat} demonstrate superior performance by integrating side features and semantic relations into the recommendation process~\cite{wang2022survey}.

The majority of publicly available Bangladeshi e-commerce datasets are constructed with a primary focus on sentiment analysis, as shown in Table~\ref{tab:dataset_comp}. Datasets proposed by \cite{sarowar2019automated}, \cite{shafin2020product}, \cite{shanto2023binary}, and \cite{rashid2024comprehensive} lack the fundamental user-item interaction records required to train any personalized recommendation model. As a result, this limitation confines their utility strictly to text classification tasks. While a limited number of studies, including \cite{shamael2024banglishrev}, \cite{hossain2021rating}, and \cite{akter2021bengali}, do contain user identifiers, authors do not demonstrate the utility of the datasets for personalized recommendation, keeping the problem formulation limited to sentiment analysis. 
Furthermore, none of the existing datasets encode explicit relational knowledge among entities (e.g., connections between products, categories, brands, and publishers) limiting their effectiveness for cold-start mitigation and explainable recommendations.

Our research bridges this gap by introducing RokomariBG. As shown in Table~\ref{tab:dataset_comp}, our dataset is uniquely designed for recommendation research, offering both essential user--item interactions and rich, structured relational knowledge in the form of a heterogeneous graph with eight relation types. Augmented with 23 side features across five entities (Book, Author, Category, Publisher, Review), RokomariBG enables the training and thorough evaluation of a comprehensive range of recommendation models, from classical collaborative filtering and factorization machines to modern neural retrieval systems, enhanced LightGCN, and HGNNs. By releasing the dataset and benchmarking, this work aims to foster future recommender systems research in Bangla e-commerce domain.

\section{Conclusion and Future Work}

In this work, we introduced RokomariBG, the first large-scale heterogeneous book graph dataset for personalized recommendation in Bangla, comprising over 127K books, 63K users, rich relational metadata, and more than 200K user reviews. Through extensive benchmarking on top-$N$ and sequential recommendation tasks, we showed that both heterogeneous relational information and textual side features play important roles in recommendation quality, highlighting challenges that are underrepresented in existing English-centric recommendation benchmarks. The dataset further exposes opportunities for studying recommendation in low-resource and code-mixed language environments, where multilingual representations may be particularly valuable. By publicly releasing the dataset, preprocessing pipelines, code, and benchmark results, we aim to provide a reproducible foundation for future research on multilingual recommendation, explainable recommendation, conversational recommendation, review understanding, and culturally grounded recommendation systems.

\section{Acknowledgements}
We thank Ahmed Wasif Reza, Atiqur Rahman, Ahmed Abdal Shafi Rasel, Nishat Tasnim, Nishat Tasnim Niloy, and Amit Mandal for helpful discussions and feedback on our earlier drafts and preliminary results during the capstone presentation.

\bibliographystyle{ACM-Reference-Format}
\bibliography{sample-base}

\appendix

\section{Qualitative Analysis}
\label{app:case_study}

To provide a concrete illustration of the model's performance, we present a case study of a sample user (\texttt{USER544691}) and \texttt{USER502420}) from the test set. 

Table~\ref{tab:sample_recommendations} compares the user's ground truth interactions (historical preferences) with the Top-10 recommendations generated by the Neural Two-Tower model.

\begin{table*}[t]
\centering
\caption{Sample Recommendation Results for Two Users}
\label{tab:sample_recommendations}
\small

\begin{tabular}{@{} p{6.5cm} p{6.5cm} @{}}
\toprule
\textbf{User ID: USER44691} & \textbf{User ID: USER02420} \\
\midrule
\textbf{Test Set (2 books):} \newline
\textit{(1)} Collection of 2 Books to Strengthen English Foundation (rating: 0) \newline
\textit{(2)} Collection of 5 Books for Primary Teacher Recruitment (rating: 0) &
\textbf{Test Set (1 book):} \newline
\textit{(1)} First Lesson of Success (rating: 0) \\
\midrule
\multicolumn{2}{c}{\textbf{Top-10 Recommendations}} \\
\midrule
1. Collection of 3 Books to Ensure Good Results in JSC \& SSC and Speak English Fluently (0.999) & 1. Stress Management (1.000) \\
2. HSC (0.998) & 2. Silent Treasure of Legend (0.999) \\
3. \textbf{Collection of 5 Books for Primary Teacher Recruitment (0.998)} \checkmark & 3. Joblit (Free Cadre Book) (0.998) \\
4. Collection of 7 Books to Get More Than 70\% Marks in University Admission Test (0.997) & 4. \textbf{First Lesson of Success (0.998)} \checkmark \\
5. IELTS Reading in Bangla (0.996) & 5. Spoken English for Beginners (0.997) \\
6. 9 Books for Science Department of Class Eleven--Twelve (0.996) & 6. Onu (0.996) \\
7. Consummate English Grammar and Composition for HSC (0.996) & 7. Road to Success (0.996) \\
8. Two Books by Monirul Islam Sir (0.995) & 8. Curse (0.996) \\
9. \textbf{Collection of 2 Books to Strengthen English Foundation (0.994)} \checkmark & 9. Bangla First Paper (Prose, Poetry, Drama \& Novel) (0.995) \\
10. Aviation Career (0.994) & 10. Prime Textile Engineering Job Solution (0.994) \\
\bottomrule
\end{tabular}

\begin{tablenotes}
\small
\item \checkmark\ denotes successfully recommended ground truth items. Values in parentheses represent model confidence scores. Both users achieved perfect hit rates, demonstrating the model's effectiveness in capturing user preferences.
\end{tablenotes}

\end{table*}

For \textbf{USER44691}, the Neural Two-Tower model successfully retrieves both ground truth items within the Top-10 recommendations, ranking \textit{\textbf{Collection of 5 Books for Primary Teacher Recruitment}} at the third position and \textit{\textbf{Collection of 2 Books to Strengthen English Foundation}} at the ninth position. For \textbf{USER02420}, the model correctly places \textit{\textbf{First Lesson of Success}} at the fourth position. These accurate rankings demonstrate the model’s effectiveness in learning meaningful user--item representations. By embedding user features and item metadata into a shared latent space, the model captures users’ preferences for educational, professional, and motivational literature.

\section{Additional Exploratory Data Analysis}
\label{app:aeda}

\paragraph{\textbf{Author productivity and popularity}} Table~\ref{tab:author-productivity} indicates that there is a wide range of difference in the author productivity across the Bangla publishing market coverd by the dataset. Rabindranath Tagore is the top ranked with 1,425 books and second in the list is Bibhutibhushan Bandyopadhyay with 868 books. This trend indicates the historical cultural role of Tagore, and also the constant reprinting of his works through the years. Moreover, beyond individual author dominance, there is also a high degree of diversity in the publishing industry, including traditional Bangla literature, modern Islamic scholars, global self-help writers, and collective editorial contributions.

\begin{table}
   \centering
   \caption{Most prolific authors by book count.}
   \label{tab:author-productivity}
   \small  %
   \begin{tabular}{clr}
      \toprule
      \textbf{Rank} & \textbf{Author Name} & \textbf{\# Books} \\
      \midrule
      1 & Rabindranath Tagore & 1,425 \\
      2 & Bibhutibhushan Bandyopadhyay & 868 \\
      3 & Maulana Ashraf Ali Thanvi & 619 \\
      4 & Mufti Muhammad Taqi Usmani & 569 \\
      5 & Rakib Hasan & 561 \\
      6 & Sunil Gangopadhyay & 552 \\
      7 & Humayun Ahmed & 506 \\
      8 & Dale Carnegie & 390 \\
      9 & Anisul Haque & 341 \\
      10 & Sarat Chandra Chattopadhyay & 284 \\
      \bottomrule
   \end{tabular}
\end{table}

\begin{table}
\centering
\caption{Top 10 authors by user interactions}
\label{tab:top-authors-interactions}
\small
\begin{tabular}{@{}clcc@{}}
\toprule
\textbf{Rank} & \textbf{Author} & \textbf{\# Books} & \textbf{\# Reviews} \\ \midrule
1 & Arif Azad & 62 & 11014 \\
2 & Humayun Ahmed & 506 & 8526 \\
3 & Munzereen Shahid & 8 & 3349 \\
4 & Saiful Islam & 12 & 4934 \\
5 & Ayman Sadiq & 27 & 2660 \\
6 & Dr. Khandkar Abdullah Jahangir & 68 & 3143 \\
7 & Dr. Mizanur Rahman Azhari & 11 & 2765 \\
8 & Sadman Sadiq & 8 & 1790 \\
9 & Jhankar Mahbub & 44 & 3010 \\
10 & Mohammad Nazim Uddin & 68 & 1879 \\ \bottomrule
\end{tabular}
\end{table}

\paragraph{\textbf{Linguistic Distribution}} Table~\ref{tab:review-classification} presents the different language types of the reviews in the dataset. The analysis reveals that Bangla (Unicode) constitutes the majority of reviews at 62.4\% (129,826 reviews), then English with 26.7\% (55,525 reviews), and finally, Bangla + English with 10.4\% (21,703 reviews). Only a small number (1.2\% or 2,543 reviews) is in the category of Other. This linguistic diversity, totaling of 209,597 reviews, reflects the multilingual nature of the Bangladeshi online book retail ecosystem, where readers express their viewpoints using different scripts and language preferences.

\begin{table}
\centering
\caption{Classification of reviews by language type}
\label{tab:review-classification}
\small
\begin{tabular}{lrr}
\toprule
\textbf{Language Type} & \textbf{\# Reviews} & \textbf{Percentage (\%)} \\
\midrule
English                & 55,525  & 26.7 \\
Bangla + English               & 21,703  & 10.4 \\
Bangla                 & 129,826 & 62.4 \\
Other                  & 2,543   & 1.2  \\
\midrule
\textbf{Total}         & \textbf{209,597}  \\
\bottomrule
\end{tabular}
\end{table}

\begin{table*}
\centering
\caption{Rating frequency distribution}
\label{tab:rating-distribution}
\small
\begin{tabular}{@{}lrrr@{}}
\toprule
Rating & Count & Percentage (\%) \\
\midrule
0 & 36,491  & 17.41 \\
1 & 7,208   & 3.44  \\
2 & 2,752   & 1.31  \\
3 & 6,955   & 3.32  \\
4 & 18,255  & 8.71  \\
5 & 137,941 & 65.81 \\
\bottomrule
\end{tabular}
\end{table*}
\paragraph{\textbf{Rating Sentiment Analysis}}
Table~\ref{tab:rating-distribution} shows a strongly 
polarized rating distribution: 65.81 \% of users gave 5 stars, reflecting high satisfaction, while 17.41 \% gave 0 stars. Ratings in the 1–3 star range are minimal, indicating users are mainly motivated to review when their experience is extremely positive or extremely negative.

\begin{table*}
\centering
\caption{Top 15 categories by user interactions}
\label{tab:rokomari_categories}
\small
\begin{tabular}{@{}clcc@{}}
\toprule
\textbf{Rank} & \textbf{Category} & \textbf{\# Books} & \textbf{\# Reviews} \\ \midrule
1 & English Grammar and Language Learning & 491 & 10905 \\
2 & Computer, Internet, Freelancing and Outsourcing & 1181 & 9501 \\
3 & Self-Help, Motivational and Meditation & 1826 & 9560 \\
4 & Language \& Dictionary & 55 & 9474 \\
5 & Contemporary Novel & 836 & 6819 \\
6 & Islamic Ideal & 318 & 7476 \\
7 & Career Development & 642 & 5586 \\
8 & Thriller & 569 & 4691 \\
9 & Sirate Rasul S.A.W & 1431 & 5378 \\
10 & Novel: Thriller \& Adventure & 565 & 4291 \\
11 & IELTS & 184 & 4840 \\
12 & Mystery, Detective, Horror, Thriller and Adventure & 1117 & 4066 \\
13 & Computer Programming & 760 & 4917 \\
14 & July Triumph!! & 1170 & 3397 \\
15 & Student Life Development & 167 & 3667 \\ \bottomrule
\end{tabular}
\end{table*}

\begin{table*}
\centering
\caption{Frequency of user interactions with respect to book length.}
\label{tab:page-review-summary}
\small
\begin{tabular}{@{}lrrr@{}}
\toprule
\textbf{Page Range} & \textbf{\# Book} & \textbf{Books (\%)} & \textbf{\# Reviews (\%)} \\
\midrule
1--100       & 25,550 & 25.47 & 15.83 \\
101--200     & 27,590 & 27.50 & 36.95 \\
201--300     & 15,507 & 15.46 & 16.38 \\
301--400     & 10,906 & 10.87 &  8.96 \\
401--500     &  6,364 &  6.34 &  4.90 \\
500+         & 14,404 & 14.36 & 16.97 \\
\bottomrule
\end{tabular}
\end{table*}

\paragraph{\textbf{Categories by user engagement}}
Table~\ref{tab:rokomari_categories} shows that \textit{English Grammar and Language Learning} dominates user engagement with 10,905 reviews, followed by computer/freelancing topics (9,501 reviews). Religious and Islamic books, self-help books, career growth books and thriller/ adventure books also have a high interaction as it shows that the user is interested in spiritual, educational and motivational books.

\paragraph{\textbf{Book Page Range Distribution}}
Table~\ref{tab:page-review-summary} reveals that mid-length books (101--200 pages) dominate reviews (36.95 \% of total), even though 27.50 \% of the collection) is represented. Short books (1--100 pages) receive relatively fewer reviews (15.83 \%)  despite comprising  (25.47 \%) of the dataset, indicating limited reader engagement. Books with higher page counts (500+ pages) attract a fair proportion of engagement (16.97 \%). Overall, readers show strongest preference for mid-length titles.

\begin{table*}
\centering
\caption{Top 20 books by number of reviews}
\label{tab:top20books-by-reviews}
\small
\begin{tabular}{@{}clcc@{}}
\toprule
\textbf{Rank} & \textbf{Book Title} & \textbf{\# Reviews} & \textbf{Avg Rating} \\ \midrule
1 & Paradoxical Sajid & 2158 & 4.66 \\
2 & Bela Furabar Age (Before noon) & 2064 & 4.73 \\
3 & Message (The Message) & 1737 & 4.71 \\
4 & Englishe Durbolder Jonyo (For The Weak In English) & 1683 & 4.76 \\
5 & Ghore Boshe Spoken English (Spoken English at Home) & 1667 & 4.36 \\
6 & Paradoxical Sajid 2 (Paradoxical Sajid 2) & 1243 & 4.60 \\
7 & Shohoj Bhashay English Grammar (English Grammar in Simple Language) & 1136 & 4.79 \\
8 & Englishe Durbolder Jonyo Vocab Therapy (Vocabulary Therapy for  Weak in English) & 1097 & 4.83 \\
9 & Freelancing  & 986 & 3.45 \\
10 & Jibon Jekhane Jemon (Where life is like) & 941 & 4.69 \\
11 & Shobar Jonyo Vocabulary (Vocabulary For Everyone) & 714 & 4.44 \\
12 & Student Hacks  & 713 & 4.33 \\
13 & Communication Hacks  & 668 & 4.45 \\
14 & Ek Nojore Quran (Quran at a Glance) & 599 & 4.70 \\
15 & Rasulullah (SAW) Er Shokal Shondhar Dua (Morning and Evening Dua and Zikr) & 595 & 4.79 \\
16 & Productive Muslim (Productive Muslim) & 585 & 4.60 \\
17 & Smart English Is The Smart Way To Learn English  & 576 & 4.02 \\
18 & Be Smart With Muhammad (SAW) & 548 & 4.52 \\
19 & Never Stop Learning  & 521 & 4.37 \\
20 & HSC Joibo Jouger Interactive Map ( Hsc Interactive Map Of Organic Compounds) & 358 & 4.55 \\ \bottomrule
\end{tabular}
\end{table*}

\begin{table*}
\centering
\caption{Publisher-to-publisher affinity via Jaccard similarity on shared authors}
\label{tab:top10-publisher-networks}
\small
\begin{tabular}{cllrr}
\toprule
\textbf{Rank} & \textbf{Publisher 1} & \textbf{Publisher 2} & \textbf{\# Shared Authors} & \textbf{Similarity} \\
\midrule
1  & Anannya Books        & Somoy Prakashan Books          & 64 & 0.178 \\
2  & Anyaprokash Books    & Kathaprokash Books             & 61 & 0.177 \\
3  & Banglaprokash Books  & Panjeree Publications Books    & 58 & 0.173 \\
4  & Mowla Brothers Books & Agamee Prokashoni Books        & 58 & 0.154 \\
5  & Anyaprokash Books    & Somoy Prakashan Books          & 57 & 0.172 \\
6  & Anyaprokash Books    & Anannya Books                  & 57 & 0.170 \\
7  & Oitijjhya Books      & Kathaprokash Books             & 53 & 0.115 \\
8  & Abosar Prokashana Sangstha Books 
   & Protik Prokashana Sangstha Books & 52 & 0.192 \\
9  & McGraw               & McGraw Hill Education (India)  & 52 & 0.161 \\
10 & Prothoma Prokashan Books & Kathaprokash Books         & 52 & 0.130 \\
\bottomrule
\end{tabular}
\end{table*}

\paragraph{\textbf{Publisher Affinity via Shared Authors}}
Table~\ref{tab:top10-publisher-networks} highlights strong ties among Bengali publishers based on shared authorship. The similarity column compares the relative strength of the relationships using the Jaccard Similarity. The result reveals a high overlap within the local publishing ecosystem, with the strongest pairs being Anannya Books-Somoy Prakashan (64 authors, 0.178 similarity), followed by Anyaprokash-Kathaprokash (61 auhtors, 0.177) and Banglaprokash-Panjeree (58 authors, 0.173). 

\paragraph{\textbf{Category Cold-Start Analysis}} In Table~\ref{tab:coldstart-by-category}, there is a great difference in the severity of cold starts (books with <3 reviews) across content domains. Categories that are specialized are extremely sparse in the dataset, including Foreign Language Books (91.6\% cold-start), Biographies, Memories \& Interviews (90.0\%), and several others above 80\%. Computer, Internet, Freelancing stand at 75.3\%. The variation of these category-specific patterns (75\%--92\% range) requires domain conscious strategies.

The overall dataset sparsity patterns represent several research challenges: cold start books, low median user activity (1 review), category-level cold start rates of 75\% to 92\% . These issues inspired our Neural method that reached NDCG@10 of 0.2368  by utilizing multi-relational information to offset the sparse user-item interactions.

\begin{table*}
\centering
\caption{Cold-start severity by category}
\label{tab:coldstart-by-category}
\small
\begin{tabular}{lrr}
\toprule
\textbf{Category} & \textbf{Cold Books} & \textbf{Cold-Start \%} \\
\midrule
Self-Help, Motivational and Meditation & 1,444 & 79.1\% \\
Mathematics, Science \& Technology     & 1,426 & 89.0\% \\
Foreign Language Books                 & 1,365 & 91.6\% \\
Translated Books                       & 1,235 & 84.9\% \\
Business, Investing \& Economics       & 1,239 & 88.5\% \\
Biographies, Memories \& Interviews    & 1,155 & 90.0\% \\
History and Tradition                  & 1,080 & 88.8\% \\
Amar Ekushe Boimela                    & 999   & 83.8\% \\
Computer, Internet, Freelancing        & 889   & 75.3\% \\
Novel                                  & 948   & 84.6\% \\
Mystery, Detective, Thriller           & 858   & 76.8\% \\
\bottomrule
\end{tabular}
\end{table*}

\section{Additional Experiments}
\label{app:experiments}

\begin{table*}
\centering
\small
\caption{Hyper-parameter search space and best configuration for neural two tower retrieval.}
\label{tab:hparam-search}
\begin{tabular}{lll}
\toprule
\textbf{Hyper-parameter} & \textbf{Search candidates} & \textbf{Best} \\
\midrule
\texttt{id\_emb\_dim} & \{64, 96, 128, 192\} & 128 \\
\texttt{text\_emb\_dim} & \{128, 192, 256, 384\} & 256 \\
\texttt{out\_dim} & \{128, 192, 256, 384\} & 256 \\
\texttt{item\_mlp\_layers} & \{2, 3, 4\} & 2 \\
\texttt{user\_mlp\_layers} & \{2, 3, 4\} & 2 \\
\texttt{item\_hidden\_dim} & \{128, 256, 384, 512\} & 256 \\
\texttt{user\_hidden\_dim} & \{128, 256, 384, 512\} & 256 \\
\texttt{dropout} & \{0.0, 0.1, 0.2\} & 0.1 \\
\texttt{batch\_size} & \{128, 256, 512, 1024\} & 256 \\
\texttt{lr} & \{1e\!-\!4, 3e\!-\!4, 5e\!-\!4, 1e\!-\!3, 2e\!-\!3\} & 5e\!-\!4 \\
\texttt{weight\_decay} & N/A & 1e\!-\!5 \\
\texttt{max\_history} & \{10, 20, 50, 100\} & 50 \\

\texttt{patience} & N/A & 4 \\
\texttt{text\_model\_name} & 
\begin{tabular}[t]{@{}l@{}}
\texttt{multilingual-e5-large-instruct},\\
\texttt{multilingual-e5-base},\\
\texttt{bangla-sentence-transformer},\\
\texttt{paraphrase-multilingual-mpnet-base-v2}
\end{tabular}
& \texttt{multilingual-e5-large-instruct} \\
\bottomrule
\end{tabular}
\end{table*}

\end{document}